\documentclass[aps, prd, twocolumn, lengthcheck, superscriptaddress, 
nofootinbib]{revtex4-1}

\usepackage{epsfig}
\usepackage[usenames]{color}
\usepackage{graphicx}
\usepackage{amsmath}
\usepackage{epstopdf}

\def\bi{\bibitem}

\def\la{\langle}\def\ra{\rangle}
\def\be{\begin{eqnarray}}\def\ee{\end{eqnarray}}
\def\lsim{\mathrel{\rlap{\lower3pt\hbox{\hskip1pt$\sim$}}
     \raise1pt\hbox{$<$}}} 
\def\gsim{\mathrel{\rlap{\lower3pt\hbox{\hskip1pt$\sim$}}
     \raise1pt\hbox{$>$}}} 
\def\del{\partial}

\def\LagnEFT{${\cal L}_{\psi\hat{\chi}\rho\omega}$}

\allowdisplaybreaks

\begin{document}

\title{From  Nuclear Matter with Quenched $g_A$ to  Compact-Star Matter \\
 with a Signal for Emergent Hidden Scale Symmetry}
\author{Mannque Rho}
\email{mannque.rho@ipht.fr}
\affiliation{Universit\'e Paris-Saclay, CNRS, CEA, Institut de Physique Th\'eorique, 91191, Gif-sur-Yvette, France }

\date{\today}

\begin{abstract}
An ``unorthodox" idea is developed that the long-standing mystery in nuclear physics of the effective axial-current coupling constant in nuclei, $g_A^{\rm eff}\approx 1$,  could be interpreted in terms of an emerging hidden scale symmetry in  dense compact-star matter. Arguments are presented using an effective field theory anchored on a renormalization-group approach to interacting baryons on the Fermi surface coupled with hidden symmetric heavy mesonic degrees of freedom that enables one to go beyond  Weinberg's nuclear effective field theory involving nucleon and pion fields only, referred hereon to as $\chi$EFT$_\pi$. Both hidden local and scale symmetries, the former involving the vector mesons $\rho$ and $\omega$  and the latter the hidden scalar meson,  a dilaton $\hat{\sigma}$ (i.e., $f_0(500)$), play the crucial role. Going beyond the density regime applicable to normal nuclear  matter $n_0$,   the notion of ``hadron-quark continuity HQC)" is brought in via the skyrmion structure of the nucleon argued to be  valid in QCD at  large $N_c$ limit and the large $N^\prime$ limit of the Grassmannian model $G/H= [O(N^\prime)/O(N^\prime-p) \times O(p)]$ where $N^\prime=4$ and $p=2$ for hidden local symmetry and the IR fixed point in QCD for $N_f \leq 3$  involving ``genuine/QCD-conformal dilaton" for hidden scale symmetry.  The connection between the quenched $g_A$ and the sound speed $v^2_{s}/c^2\approx 1/3$ inside dense compact stars could be interpreted as a signal for emergent ``pseudo-conformal" symmetry

\end{abstract}

\maketitle

\setcounter{footnote}{0}
\renewcommand{\thefootnote}{\arabic{footnote}}
\vskip 1.0cm

\subsection*{\it Preamble}
%
{\it In this note I revisit the projects initiated in early 2000 with the visits to Korea Institute for Advanced Studies (KIAS) and Seoul National University (SNU). The projects consisted of two seemingly unrelated issues: One, why does the axial coupling constant $g_A$ come out so close to 1 in nuclei first seriously considered by Denys Wilkinson in his Les Houghes Lectures~\cite{wilkinson} whereas there was no obvoius reason why it should be so, contrary to the iso-vector-vector coupling constant $g_V$, and the other issue, what is the most extreme state of matter stable under gravitational collapse? 
 
These questions were raised  in my visits to KIAS with Gerry Brown, who was at the time working with Hans Bethe on supernovae, participating in  workshops, and  continued with the  5-year World-Class University (WCU) Project at Hanyang University supported by the Korean Government. The WCU project continued further at the newly established Institute for Basic Science (ISB) endowed with an RI beam accelerator for nuclear physics. These projects were followed by the collaboration with the Theory Institute in Jilin University, Changchun, China. 

What's discussed in this note is what emerged after many years of series of pros and cons for a {\bf coarse-grained} description of going bottom-up hadron-quark continuity (HQC) from normal nuclear matter to superdense matter in compact stars to match the putative top-down quark-hadron continuity (QHC) from perturbative QCD.}

\section*{From Nuclei to Compact-Star Matter}
The recent upsurge of developments in tools for astrophysical observables triggered by gravitational waves and associated methods  sharpened the necessity for trustful and reliable equations of state (EoS) of dense baryonic matter going beyond the density of equilibrium nuclear matter $n_0\sim 0.16$ fm$^{-3}$. In the past,  up to 1990s, the structure of finite nuclei and nuclear matter has been approached by ``phenomenological" models -- which could be phrased in terms of ``phenomenological" Lagrangians -- guided by observed hadronic degrees of freedom as fields with associated  or assumed symmetries involved. The approach, standardly adopted then, consisted  of resorting to fitting the experimentally available data with the arbitrary set of parameters and compute the EoS of the baryonic matter with a variety of many-body techniques to take into account for the strong correlations between baryons.  With a large number of parameters with highly sophisticated interaction Hamiltonians --  such as the Paris potential -- it was at least plausible to arrive a tenable EoS valid near $n_0$.

The arrival of gravity-wave and related observations required going beyond $n_0$ to $\sim (2-3)n_0$ making the old ``phenomenological" models incapable to address the observations.   Given that at high scales, say, at high density (or high energy), quarks, the QCD degrees of freedom, should become relevant degrees of freedom.  Hence at some high density, baryons must turn ``quarkish."\footnote{I use the terminology ``quarkish" instead of ``quark"  as will become clarified.} There must therefore be a change-over from hadrons to quarkish objects that I will refer to as quasi-quarks. It is here  that $\chi$ effective field theory ($\chi$EFT) could offer a possible way-out to the problem. This changeover of degrees freedom is referred to as ``hadron-quark continuity (HQC)." It could be a phase transition in the Ginzburg-Landau-Wilson (GLW) sense or could be a non-GLW involving quasiquarks, not genuine quarks, at the relevant density. In what follows, I adhere to the latter for the process taking place  in stars. 

The notion of EFT that I adopt for the problem is motivated by two developments: One a long-standing mystery of quenched $g_A$ in nuclear matter dating way back to 1970s~\cite{wilkinson}  and the other in 1990s a development anchored on  the line that Weinberg phrased in terms of  a ``Folk Theorem (FT)"~\cite{FT}. I suggest how the FT works out can be best illustrated in nuclear processes at low-energy scale~\cite{FT-MR}. The relevant phenomenon invoked was what I called in \cite{FT-MR}, lacking a more appropriate name, a ``chiral filtering mechanism." This, I have proposed, is  a beautiful nuclear-physics proof of the FT. 

The concept leading to the Weinberg's FT is succinctly captured in \cite{PL} entitled  ``Phenomenological Lagrangians." What's treated there is a chiral  Lagrangian with the pseudo-Nambu-Goldstone field $\pi$. It can be considered  as an ``effective" Lagrangian that results when all other degrees of freedom than $\pi$ -- including the nucleon fields figuring in nuclear physics --  are ``integrated out" from a phenomenological Lagrangian of the Paris-potential-type. Now the resulting Lagrangian must be chirally symmetric, hence the integrated-out degrees of freedom in the phenomenological Lagrangian, vector and scale mesons etc., must have been coupled to the pions in (hidden) local and  scale-symmetric way so as to lead to the resulting chirally symmetric  ``Phenomenological Lagrangian" of \cite{PL}.\footnote{The nucleon field of course must figure in nuclear physics. In my view, it can be properly considered emerging as a skyrmion.}  By now, the FT is formulated in great detail as a derivative or power expansion  in nuclear physics in what I will identify for reasons which will be clarified as $\chi$EFT$_\pi$ (where the subscript $\pi$ represents the sole bosonic field figuring together with the nucleons).  This approach is being commonly referred to, in nuclear physics community, as {\bf the} ``first-principles approach" to nuclear physics.  What I would call a ``work of the art" result  in such  approach  is nicely reviewed in \cite{bira-review} and up-dated in the most recent article \cite{Alp-etal}. There are numerous publications of similar quality in the literature that I opt not to get into. 

In doing $\chi$EFT$_\pi$, the cutoff scale is taken typically $\Lambda_{\chi EFT} \sim 3 m_\pi$. In nuclear physics, the cutoff scale cannot be precisely defined given the multiple close-lying scales involved in nuclear dynamics. The $\chi$EFT$_\pi$ with the cutoff $\Lambda_{\chi EFT}\sim 3 m_\pi$ implies the degrees of freedom heavier than the lowest-lying vectors ${\bf V}=\rho, \omega$,  the iso-scalar scalar of mass $\sim 500$ MeV identifiable with $f_0 (500)$, and possibly iso-vector scalars have been integrated out. This also means that the nucleon with mass $\sim 7 m_\pi$ must be treated in ``heavy-fermion" formalism to deal with kinematics compatible with ``soft modes" interacting with pions as is the case in low-energy nuclear physics. The mass difference $m_\Delta-m_N$ where $\Delta$ stands for the $(3,3)$ resonance is commonly taken into account. The strategy adopted in this article is that as density goes beyond  that appropriate for $\chi$EFT$_\pi$, the nucleon can also be integrated out to lead to a skyrmion together with the heavy modes.

At present there is no  reliable way to connect the $\chi$EFT$_\pi$ to  what could be taking place above the cutoff scale relevant for dense compact-star structure.  There is no known way to go from $\chi$EFT$_\pi$  over the HQC from $\Lambda_{\chi EFT}$ in a way sufficiently consistent with QCD. There are in the literature numerous model-dependent descriptions -- with or without quarkish degrees of freedom --  at a HQC density $n_{HQC}\sim (2-3) n_0$.   The approach I will adopt is based on certain features of the skyrmion description applied to  the structure of baryonic matter at higher density than $n_0$. The power of skyrmions in all areas of physics going beyond nuclear physics is illustrated in the volume~\cite{multifacet}. The key strategy I adopt relies on that in the large $N_c$ limit, the nucleon (proton, neutron) can be described by the soliton skyrmion, mapping onto the constituent quark model, and for large density, the nucleon in the matter can be reliably described in terms of skyrmion {\it crystal} structure~\cite{multifacet}. This suggests that as density increases beyond $n_{HQC}$, certain skyrmion topological structure not visible in free space is to emerge in the EoS~\cite{park-vento}.  This means that as $\chi$EFT$_\pi$ goes over the $n_{HQC}$, the properties encoded in the large $N_c$ limit in the skyrmion description would take over. 

Then the question arises as to what the relevant skyrmion Lagrangian is. The original Skyrme model -- with the Skyrme Lagrangian denoted hereon as ${\cal L}_{\rm skyrme}$ consisting of the current algebra term and the Skyrme quartic term (with the coefficient $1/g^2$) that plays a key role for stabilizing the soliton,  recently highlighted in nuclear physics in a volume by Manton~\cite{Manton} is found to work fairly -- and even surprisingly -- well in finite nuclei.  As an {\it effective} model of QCD, there is nothing however to indicate what else, apart from the current algebra term,  should or can figure in the effective Lagrangian. In fact the Skyrme quartic term  must be encapsulating a lot more than just the stabilizer of the soliton. To give an example, the binding energy of light nuclei comes out much too  large, reflecting the large $N_c$ dependence of the Skyrme Lagrangian. This problem is however significantly ameliorated by including the vector-meson terms.   

The principal theme of this article is that to address the highly dense matter with $n\gsim (2-3) n_0$ at low temperature, various ``hidden symmetry" degrees of freedom must be required in the skyrmion Lagrangian.  This requirement becomes evident when  the baryonic matter is described in terms of a skyrmion crystal. Minimally required   in  $\chi$EFT$_\pi$ Lagrangian are the hidden local vectors $V_\mu (\rho_\mu, \omega_\mu)$ and the dilaton scalar $\hat\chi=f_{\hat\chi} e^{\hat\sigma/f_{\hat\chi}}$.
\footnote{Here $\hat\chi$ stands for the ``conformal compensator field" of scale dimension 1,  not to be confused with chiral symmetry and $\hat\sigma$ for the dilaon, not to be confused with the $\sigma$ of the sigma model.} Thus the task is to make $\chi$EFT$_\pi$ go, in as simple a way  and as consistently as possible with the hidden symmetries involved, over the transition density  $n_{HQC}\gsim (2-3) n_0$  toward  compact-star densities. In \cite{MR-towards}, this task is explained in thirteen -- admittedly inelegant --   Propositions. In this note, I will minimize the number of  the Propositions eliminating redundancies,  correcting incorrect items and  injecting newer ideas.

\section*{Nuclear Effective Field Theories with Hidden-Symmeterized Chiral Lagrangian}
As mentioned, crossing over the HQC point is rendered feasible in terms of  certain skyrmion structure of nucleons encoded in QCD.  This is done by exploiting  large $N_c$ properties  of the degrees of freedom in QCD.  To do this, one incorporates -- in consistency with the symmetries involved --   both the HLS vectors $V_\mu (\rho_\mu, \omega_\mu)$ and the HSS scalar dilaton $\hat{\chi}$, into  the Lagrangian $\chi$EFT$_\pi$ to arrive at a ``generalized EFT,"  dubbed in the absence of more appropriate acronym as, G$n$EFT.  How to do this has been known since 1990's. What was achieved  recently was to provide a {\t stronger support} to what was done in \cite{MR-towards}. This involves validating the HLS Lagrangian~\cite{HLS} in a Grassmannian model~\cite{yamawaki} and the HSS  Lagrangian with a  ``genuine dilaton (GD)"~\cite{GD} and/or a  ``QCD-conformal dilaton (QCD-CD)"~\cite{Zwicky} . 


The HLS Lagrangian, when coupled to baryon fields, has been successfully applied to nuclear processes. The most important parameter of the Lagrangian ``$a=2$", playing the key role in phenomenology such as vector-meson dominance, $\rho$ universality, KSRF relation, vector manifestation etc. associated with local gauge symmetry which had remained, to a certain extent,  completely arbitrary is, most remarkably,  determined by the gauge structure of the HLS   by the large $N^\prime$  limit of the Grassmannian model $G/H= [O(N^\prime)/O(N^\prime-p) \times O(p)]$ for $N^\prime=4$ and  $p=3$~\cite{yamawaki}.\footnote{Briefly stated,  it gives $G/H=O(4)/O(3)\simeq SU(2)_L\times SU(2)_R/SU(2)_V$ with the $\rho$ meson being $[SU(2)_V]_{local}$ acquiring the kinetic energy term in the HLS of \cite{HLS} as $N^\prime\to \infty$ is taken.  What is noteworthy is that the ``a=2" results come from pure dynamics at the quantum level at large $N$ limit.  Furthermore what's called the chiral cutoff scale $\sim 4\pi f_\pi$ can be regarded as a matching scale of the HLS  as a ``magnetic dual" to QCD. The unbroken phase with a massless $\rho$ meson may be realized as a {\it novel} chiral-restored hadronic phase in the hot/dense matter, not what has been searched in heavy-ion dilepton production experiments.}

How to implement HSS in nuclear dynamics is much less straightforward because the conformal structure in QCD for the flavor number $N_f\sim (2-3)$ is still highly controversial even though it has been around since 1970s.  We will follow the most recent development~\cite{GD,Zwicky,Zwicky-pion} that posits an IR fixed point in the QCD sector for $N_f\leq 3$ relevant to nuclear and astrophysics. The fixed-point is characterized by that it  accommodates the Nambu-Goldstone bosons $\pi$ and $\hat{\sigma}$ with however non-zero $f_\pi$ and $f_{\hat\chi}$, hence with massive matter fields (such as vector mesons,  nucleons and hyperons) present.  In both GD and QCD-CD, the strangeness is considered massive, so the s-quark mass is taken into account in the Lagrangian. How to incorporate scale symmetry (or ``HSS-ize")  in the HLS Lagrangian is to use the scale-dimension-1 conformal compensator field   $\hat{\chi}$.  There are some differences between GD and QCD-CD in constructing the hidden-symmetrized Lagrangian but to the leading order in the combined chiral-scale power counting, $O(\partial^2)\sim O(p^2) \sim O(m_\pi^2)\sim O(\delta)$
where $\delta=(\alpha_s - \alpha_{\rm IR})$,  they give more or less the same results~\cite{sHLS}. To higher orders, they are highly unwieldy with uncontrollable number of parameters and cannot be profitably applied as in $\chi$EFT$_\pi$.  Fortunately, however,  there is a systematic procedure to do higher-order calculations.
\section*{Hadron-Quark Continuity}
How to overcome the difficulty in doing an EFT calculation at higher density going above $\chi$EFT$_\pi$, first suggested in \cite{MR91},  was to formulate strong baryon correlations on the Fermi surface,  leading to Landau(Migdal) Fermi-liquid theory~\cite{migdal}\footnote{In this paper, Landau Fermi-liquid should stand for  Landau-Migdal Fermi liquid given that pions play a key role in nuclear physics. This distinction will not be made in what follows.}. The idea was not totally novel then. It was to exploit the four-fermi interactions given in $\chi$EFT$_\pi$, marginal in the chiral-scale counting,  as Fermi-liquid fixed point terms in the chiral expansion~\cite{gelmini,doubledecimation}~and taking them as Landau interaction parameters, do the Walecka relativistic mean-field-type calculation leading to Fermi-liquid description of  many-body correlations~\cite{matsui}.  This chain of reasonings, present thus far in a disconnected fashion, was transformed into an EFT with systematic power counting  by the strategy of an RG approach to interacting quasiparticles on the Fermi surface first developed in condensed matter physics~\cite{shankar-polchinski} and then applied in nuclear physics in \cite{friman-rho}.   The hadronic system is a lot more complicated  than condensed matter systems, so one would expect the results to be highly challenging.  It turns out however to come out surprisingly (perhaps too ?)  simple.

In formulating the continuity from hadrons to quarks (HQC), the power counting in $\chi$EFT$_\pi$ ceases to be effective. For this, the large $N_c$ property of the skyrmion crystalline structure of the nucleon assumed to be in QCD proper~\cite{park-vento} plays the crucial role.  
\section*{Generalized Nuclear EFT (G$n$EFT)}
The chiral Lagrangian ``HS-ized" with HLS and HSS,  with the topological properties invoked below taken into account, denoted here on as ${\cal L}_{\psi\hat{\chi}\rho\omega}$,  can be mapped via the Shankar-Polchinski formalism~\cite{shankar-polchinski} to a nuclear effective field theory of Landau Fermi-liquid (LFL) approach suggested in \cite{MR91,friman-rho}. Here the mean-fields of ${\cal L}_{\psi\hat{\chi}\rho\omega}$ are taken as the Fermi-liquid  fixed point quantities, i.e., Landau masses, Landau interaction parameters etc.,  in developing a systematic $1/\bar{N}$ expansion going beyond the LFL fixed point replacing the  chiral expansion in  $\chi$EFT$_\pi$.  The expansion replacing  Weinberg's chiral expansion in $\del^2\sim p^2\sim m_\pi^2$  is made in terms of the power of  $1/\bar{N}$ where $\bar{N}=k_F/({\Lambda_{\rm FS}-k_F})$ with $\Lambda_{\rm FS}$ the cutoff on top of the Fermi surface measured from the origin of the Fermi sphere~\cite{doubledecimation}. 

In what follows, this approach to nuclear EFT concisely sketched above  will be broadly referred to as G$n$EFT. 

\subsubsection*{\it Cusp in $E_{\rm sym}$}
Let me start by illustrating the principal points involved in exploiting the topology structure. The distinguished topological feature of the skyrmion crystalline structure is that as the crystal size is decreased, a skyrmion fractionizes into two 1/2-skyrmions with the half-skyrmions  bound by the monopole, so that although the chiral condensate $\bar{q}q$ is zero space-averaged, the pion decay constant is not equal to zero. This resembles the pseudo-gap phase in superconductivity. The chiral symmetry is not restored, hence there is no phase transition to the Wigner-Weyl phase in the Wilson-Ginzburg-Landau sense. There the skyrmion-to-half-skyrmion changeover taking place at a density which is not determined by theory but expected at $n_{1/2}\sim (2-3) n_0$ must involve a topology change.    

\begin{figure}[h]\centering 
\includegraphics[width=5cm]{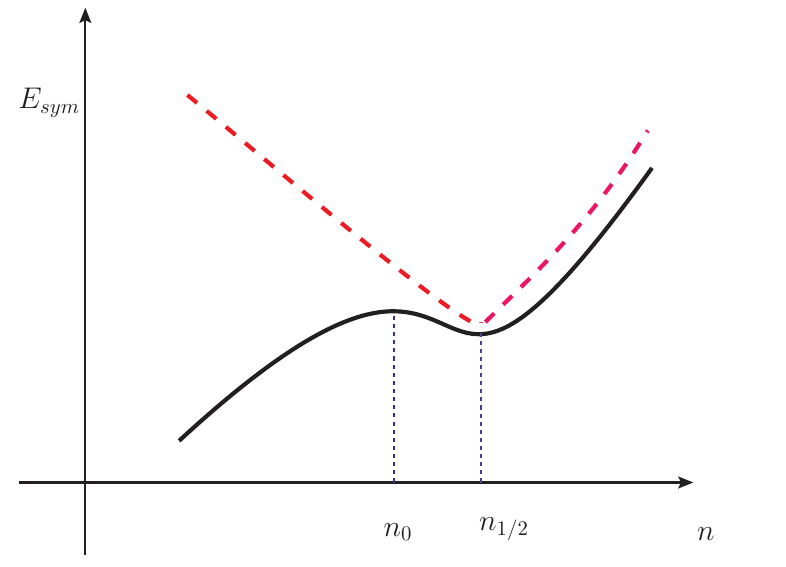}\includegraphics[width=4.cm]{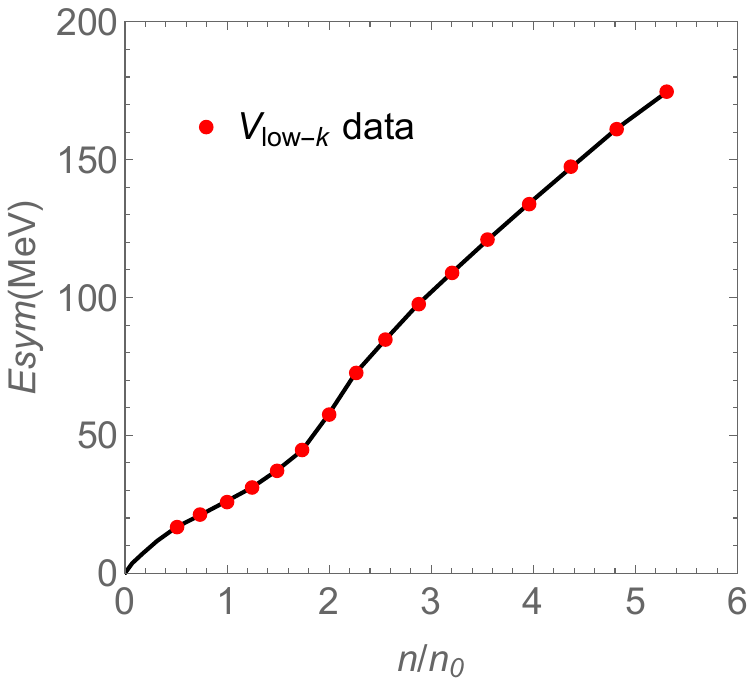}
\caption{Left Panel:  Schematic illustration of the cusp in the symmetry energy $E_{sym} (n)$ in  skyrmion crystal (dashed line) and smoothed by the interfernce between the current-algebra and the Skyrme quartic term (solid line). Right panel: $V_{\rm lowk}$ RG treatment of Landau fixed-point in G$n$EFT dominated by the nuclear tensor forces~\cite{paengetal}.}
\label{LPR}
\end{figure}

The first important property of skyrmion matter in the topology change, unexpected in going to high density in $\chi$EFT$_\pi$, is the cusp structure in the symmetry energy $E_{sym}$ in the equation of state (EoS). It is generated by the emergence of half-skyrmions at $n_{1/2}$.\footnote{Half-skyrmions also figure in condensed matter systems in lower dimensions, but they are deconfined~\cite{senthil}, whereas in hadronic matter, they are most likely confined by monopoles~\cite{cho-monopole}.}   There the dilaton condensate  $\la\hat{\chi}\ra$ sliding in density in nuclear medium driven by the vacuum change ~\cite{BRscaling} plays an important role in the structure above $n_{1/2}$.\footnote{What's called ``BR scaling" in the literature has been totally misunderstood in heavy-ion community. As already noted in \cite{BRscaling},  it can be related to the scaling of $f_\pi$, i.e., chiral condensate,  at low density (and also  temperature) but not at high density (or temperature).  This feature becomes important in the compact-star physics as well as in the quenching of $g_A$ in nuclear medium.} This has been noted early in the development~\cite{lee-park-rho} in the  Skyrme Lagrangian ``HS-zed" by the  conformally compensated dilaton field $\hat\chi$,   ${\cal L}_{\hat{\chi}{\rm Skyrme}}$. In Fig.\ref{LPR}, in the left panel, is shown  the cusp structure of the symmetry energy $E_{\rm sym}$ in the skyrmion crystal  appearing at $O(1/N_c)$ of the energy density (dashed-line), which is smoothed by the interference between the current-algebra term and the Skyrme quartic term (solid line).  This feature can be reproduced in standard nuclear many-body approaches by taking into account the BR-scaling $\rho$-meson mass and the behavior of the tensor forces affected by the cancellation of tensor strength at different nuclear densities~\cite{lee-park-rho}.\footnote{This indicates the skyrmion structure of the nucleons is implicitly involved in many-body approaches with symmetries correctly implemented.}  The dilaton condensate can affect the symmetry energy for $n\gsim n_{1/2}$ but it leaves the density $n_{1/2}$ unchanged. The right panel in the figure gives the results of the $V_{\rm lowk}$ RG formalism to confront with experiments. The cusp structure accounts for an additional repulsion brought in at $n_{1/2}$, followed by softening, not present in $\chi$EFT$_\pi$.\footnote{I will mention in the last section that this first-hardening followed by softening is present in the top-down approach ``quark-hadron continuity."} 

Here let me make a brief  remark on a point stressed by Yamawaki~\cite{yamawaki}, namely that the cusp  {\it embodies} the effect of the intricate kinetic energy term of the Grassmannian model that encodes the  complete effects of the Skyrme quartic term stabilizing the soliton and giving the full ``$a=2$ results" of the HLS Lagrangian~\cite{HLS}. 
\subsubsection*{\it Pseudo-Conformality: Manifestation in Crystal Lattice}
It is worth stressing here the most striking property generated by the half-skyrmions which is the onset of what's referred to as ``pseudo-conformality"\footnote{This may be a misnomer but its meaning will be made clearer later.}. It is what gives rise to the sound speed $(v_{s}/c)^2 \sim 1/3$ in the EoS although conformal symmetry is not restored, hence not what corresponds to the conformal velocity $(v_{conf}/c)^2 =1/3$ in the chiral limit. The potential corrections making  $(v_{s}/c)^2$ deviate from 1/3 could eventually  be calculated by $1/\bar{N}$ corrections via the coadjoint orbits~\cite{coadjoint}, a formalism that is  being developed in condensed matter physics.

\begin{figure}[h]
\begin{center}
\includegraphics[width=4.0cm]{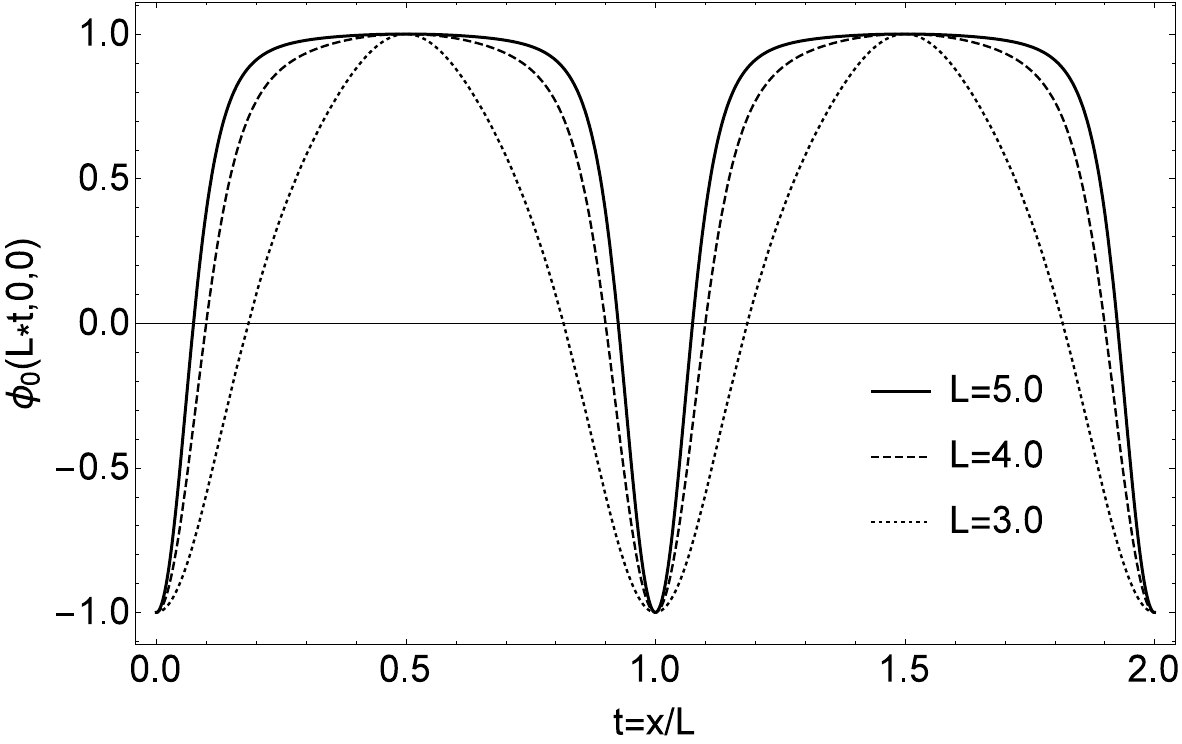}\includegraphics[width=4.0cm]{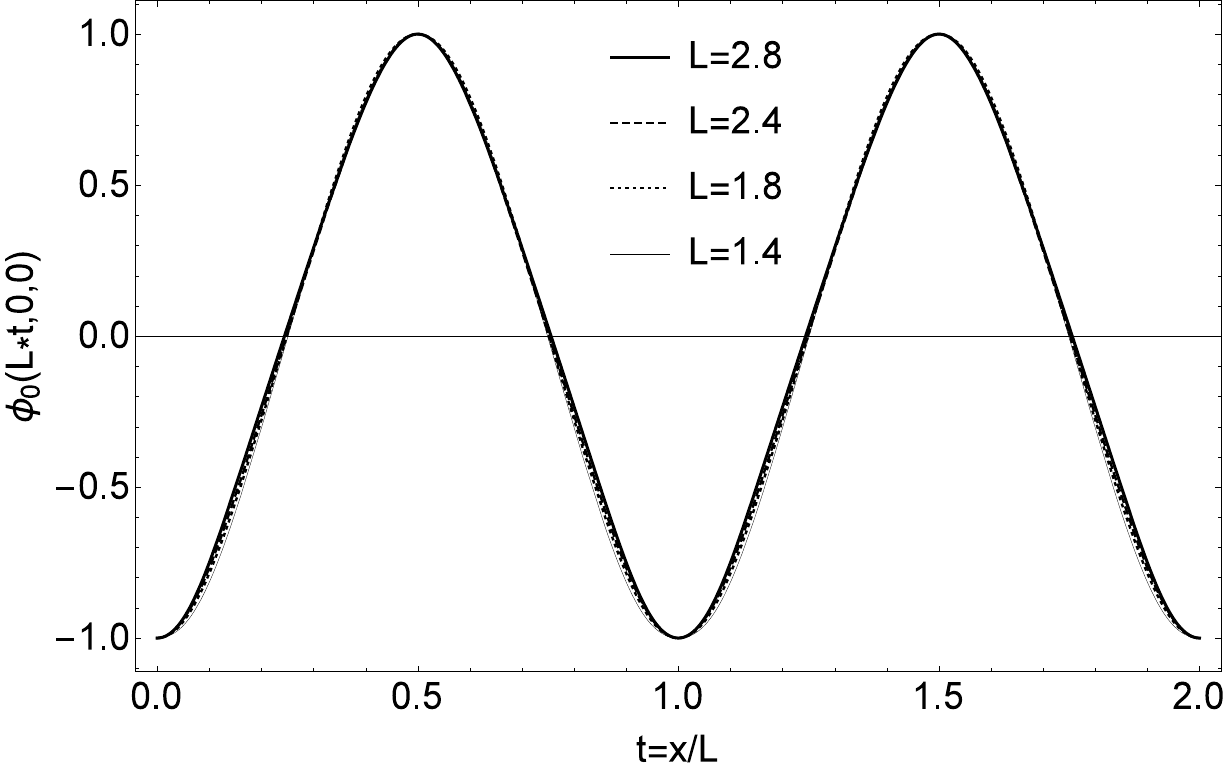}
\includegraphics[width=4.0cm]{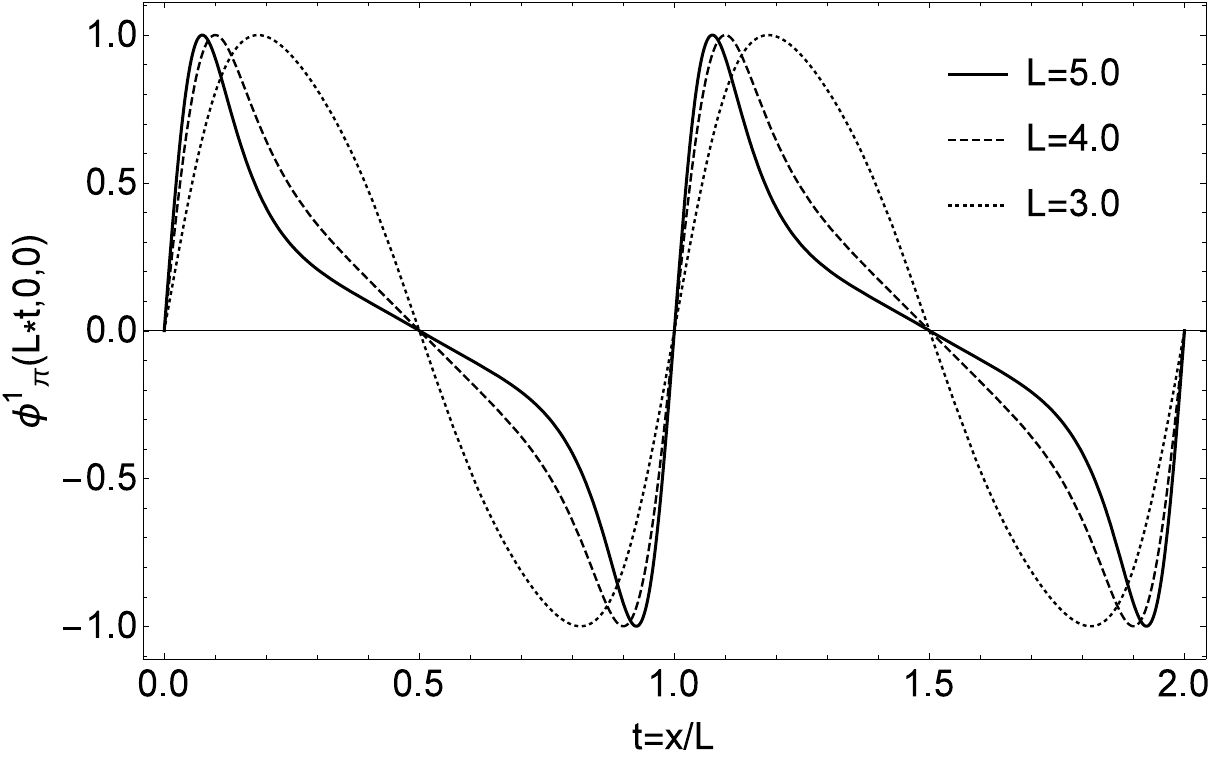}\includegraphics[width=4.0cm]{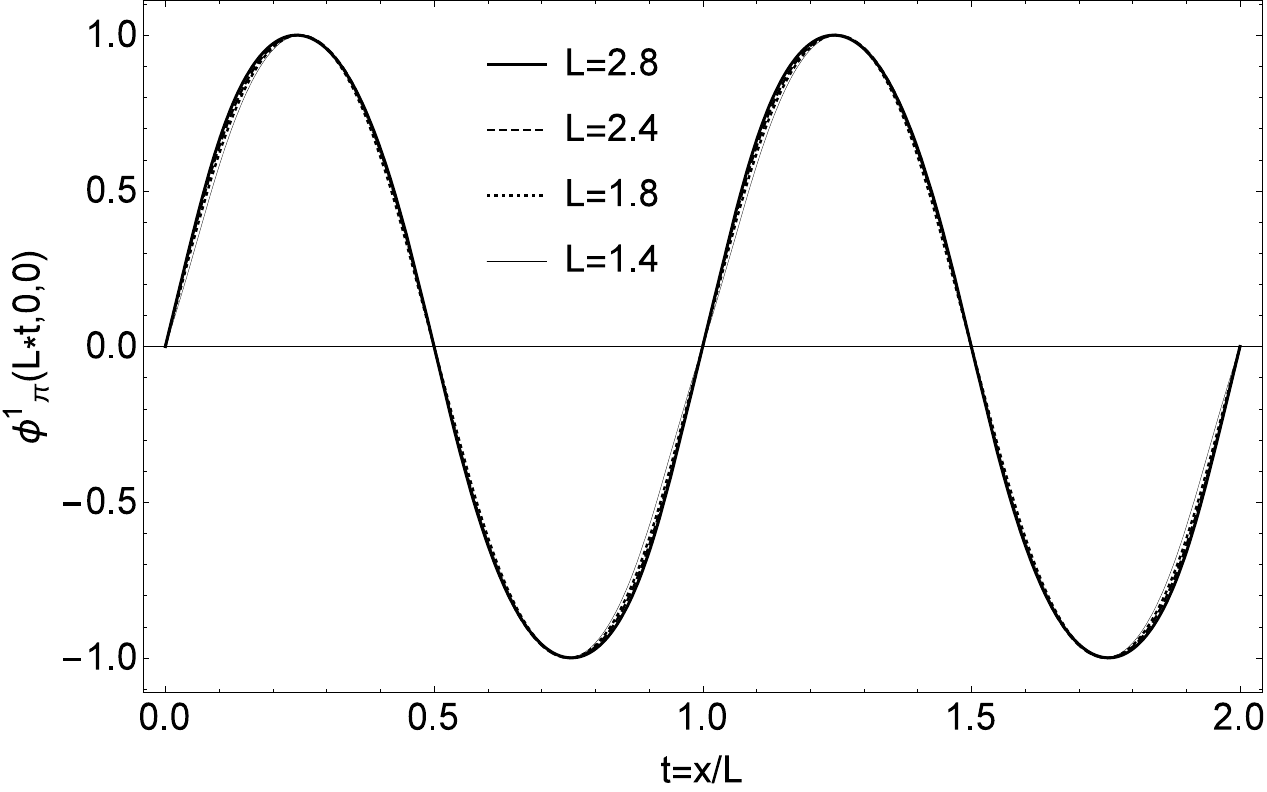}
\caption{The field configurations $\phi_0$ and $\phi^1_\pi$  as a function of $t = x/L$ along the y = z = 0 line. The maximum values for $\eta=0,\pi$ are $\phi_{0,\,L,\, {\rm max}} = \phi_{\pi,\,L,\, {\rm max}} = 1$. The left panel represents the skyrmion phase and the right panel the half-skyrmion phase that  sets in when $L\lsim L_{1/2}$. }\label{scale_inv}
 \end{center}
\end{figure}

The onset of pseudo-conformality can be illustrated~\cite{paengetal} by considering the reparametrized field $\Sigma (\vec{x}) \equiv \hat{\chi}(\vec{x}) U(\vec{x})$ where $U(\vec{x})$ is the usual chiral field $U=e^{i\pi/2 f_\pi}$.  Write $\Sigma$ as $\Sigma(\vec{x}) = \phi_0(\vec{x}) + i \phi^j_\pi(\vec{x}  )\tau^j$  with $j=1,2,3$.  Including $\rho$ and $\omega$, write the fields placed in the lattice size $L$ as $\phi_{\eta,\, L}(\vec{x}\,)$ with $\eta =0,\, \pi,\, \rho,\, \omega$ and normalize them with respect to their maximum values denoted $\phi_{\eta,L,{\rm max}}$ for given $L$.  With the HS-ized HLS treated in the mean-field  which corresponds to the Fermi-liquid fixed point,  it can be shown as in \cite{atiyah-manton} that in the half-skyrmion phase  with $L\lsim L_{1/2}$, the field configurations are found to be invariant under scaling in density as the lattice is scaled from $L_1$ to $L_2$. 
\be
\frac{\phi_{\eta,\,L_1}(L_1\vec{t}\,)}{\phi_{\eta,\,L_1,\,{\rm max}}}= \frac{\phi_{\eta,\,L_2}(L_2\vec{t}\,)}{\phi_{\eta,\,L_2,\,{\rm max}}}.
\ee %
Since other fields are quite similar, what's shown  in  Fig.~\ref{scale_inv} is the case of $\phi_{0,\pi}$ for  $\phi_{0,\pi} (t, 0,0)$ vs. $t$ with $t\equiv x/L$.  One sees  that density-scale invariance sets in for $L\lsim L_{1/2}$: The field is independent of density in the half-skyrmion phase with $L\lsim L_{1/2}$  whereas for the skyrmion phase with lower density with $L >  L_{1/2}$, it is appreciably dependent on density.
This feature will be the signal for what we call ``pseudo-conformality."

\vskip 0.2cm
 $\bullet$ {\bf Dilaton-limit fixed point}:
 
 The first important observation to make is what happens when the limit on the mean-field of $\Sigma\to 0$ in \LagnEFT is taken.  This is equivalent to doing the Fermi-liquid fixed point approximation in G$n$EFT.  The most important feature for both $g_A$ and sound speed problem we will come to later is that at the fixed point, (1) the $\rho$ mesons decouple from the Nambu-Goldstone fields and other matter fields and (2) gives rise to two terms singular in the limit~\cite{paengetal}. The  elimination of these singular terms gives what is referred to as ``dilaton-limit-fixed-point (DLFP)" constraints
\be
g_V\to g_A\to 1, \  f_\pi\to f_{\bar{\chi}}\neq 0.\label{DLFP}
\ee
Here $g_{V,A}$ correspond to the renormalized  {\it quasiparticle} vector and axial-vector coupling constants in G$n$EFT arrived at  high density  near the restoration of scale symmetry. That $g_V \to 1$ is no surprise, the CVC,  but that $g_A\to 1$ could be associated with an emerging symmetry as I will suggest. The DLFP density is given, up-to-date,  neither by theory nor by experiment.  It is however clear that they cannot be identified as genuine ``quarkish" objects. This is because  the flavor $U(2)$ symmetry of the HLS bosons breaks down with the isovector vector mesons {\it decoupling before reaching the DLFP} while the $\omega$ remains coupled and massive. The constraints (\ref{DLFP}) support two important predictions of the theory:
\begin{enumerate}
\item $g_A\to 1$ as density approaches $n\approx n_{DLFP}$.  This means that the effective quasiparticle $g_A^{\rm eff} \approx 1$ in baryonic matter at {\it some} high density. At what density this sets in will be addressed later in connection with the proposed pseudo-conformal sound speed in compact stars.
\item $f_\pi\approx f_{\bar\chi} \neq 0$ at $n\approx n_{DLFP}$.    This feature which is drastically different from conformal window~\cite{appelquist}  is consistent with the notion of an IR fixed-point in the GD/QCD-CD scenario~\cite{GD,Zwicky}. This fixed point accommodates not only the Nambu-Goldstone bosons $\pi$ and $\hat{\sigma}$ but also massive hadrons such as nucleons, vector mesons etc.
\item One striking ``unorthodox" point which will figure importantly for the sound velocity in this G$n$EFT formulation is the emergence of parity doubling in nucleons with the quasiparticle mass $m^\ast_N\to m_0\propto {\rm const.}$ in the half-skyrmion phase as $n\to n_{DLFP}$~\cite{HKL}.  The parity doubling  is not  present {\it ab initio} in zero-density medium at variance with the models that assume the $m_0$ term in QCD. This property results from an intricate interplay between the $\omega$ (coupling strength) and $f_{\hat\chi}\to m_0$ (to be explained later) with {the $\rho$ mesons} decoupled. This is the mechanism that  leads to the pseudo-conformality where the VEV of the trace of the energy-momentum tensor is non-zero but density independent. 
\end{enumerate}

$\bullet$ {\bf Nuclear Electro-Weak Response Functions}:

Another important feature of the G$n$EFT approach  is the low-energy theorems that follow from the EFT. Here I give a precision test of G$n$EFT in EM responses in nuclei at low energy below the HQC density.  It illustrates how the G$n$EFT approach fares in the regime where $\chi$EFT$_\pi$ works well.

In dealing with pseudo-conformality in baryonic matter, there is a subtle difference between the GD and the QCD-CD as to how to incorporate the scale anomaly. They are treated differently in the two references but for my purpose the difference can be avoided.  Let me first take the QCD-CD way of handling  $\beta^\prime_{\rm IR}$, the slope of the $\beta$ function at the IR fixed point~\cite{Zwicky} and discuss later what the GD scenario gives. 

Let me first assume, following the argument of  \cite{Zwicky}, that
\be
\beta^\prime_{\rm IR}=0.
\ee
Then in the $\omega\to 0, q/\omega\to 0$ limit appropriate for the LFL approach, and adopting the non-relativistic flavor $SU(6)$  symmetry\footnote{One can use relativistic formulas, but relativity does not bring appreciable differences.}, the results obtained in \cite{friman-rho} are reproduced exactly by G$n$EFT. So I will just quote them here. To do this let me first cite a few quantities most essential for nuclear interactions needed in addition to what's in condensed-matter systems. They are the  Brown-Rho (BR) scaling factor $\Phi (n)=\la\hat{\chi}\ra_n/\la\hat{\chi}\ra_0$~\cite{BRscaling} that governs how physical quantities scale in density as the vacuum changes in medium and the pionic Landau parameter $F_l^\pi (n)$ coming from the Fock term\footnote{Landau parameters in other channels will not figure in what I will discuss, so I won't list them here. See \cite{friman-rho} for them.}.  The first account for scale symmetry and the second for chiral symmetry in nuclear matter.

While my interest is focused  on the combined scale and chiral symmetries, the chiral filter mechanism~\cite{MR91,KDR} in EM responses has a direct implication for the validity of G$n$EFT at low energy and density. The quantities most interesting  are the quasiparticle gyromagnetic ratios  $\delta g_l$ in nuclei. Since the nucleon mass must scale in dense medium with the BR scaling $\Phi(n)$, one would naively think that  $U(1)_{EM}$ would be violated given that the gauge invariance would require the iso-scalar quasiparticle $\delta g_l^s$ should be $\propto 1/m_N$, not $1/m_N^\ast$ with the scaling factor $\Phi (n)$. This is nicely remedied by the back-flow diagram required by the chiral filter mechanism~\cite{KDR}. 

Even more interesting is the iso-vector quasiparticle gyromagnetic ratio given by the Migdal formula~\cite{migdal} $\delta g_l^{proton}$.  As far as I am aware, this quantity comes out wrong in  $\chi$EFT$_\pi$ approaches ~\cite{HRW}.  The prediction in G$n$EFT involves three Fermi-liquid fixed point quantities,  ${F}_1^\pi$, the Landau mass $m_L$ and the BR scaling $\Phi$.  It comes out to be  $\delta g_l^{proton}=\frac 49[1/\Phi -1-\frac 12 \frac {m_N}{m_L}  {F}_1^\pi]\tau_3$.  This agrees beautifully with the available experiments~\cite{friman-rho}.  This illustrates the extreme simplicity involved at $q/\omega\to 0$ in EM response functions given in the Fermi-liquid fixed-point approximation.

The response to the weak field is  a bit more involved but still quite simple. It is this that is related to the pseudo-conformal sound speed in stars and makes a surprising prediction that links to dense matter EoS relevant for compact-star physics. It involves the same Landau parameters $m_L$ and $\Phi$ in a subtle interplay in the in-medium properties of  the dilaton $\hat{\chi}$ and the iso-scalar vector meson $\omega$ at high density above the HQC regime.  What turns out to play a more subtle role than in the EM case  is the notion of the large $N_c$ limit  in the Goldberger-Treiman relation~\cite{NRZ}. The result obtained in \cite{friman-rho}
\be
m_L/m_N=\Phi\sqrt{\frac {g_A^L}{g_A}}\label{GT}
\ee 
yields the relation
\be
g_A^L=g_A (1-\frac 13 \Phi \tilde{F}_1^\pi)^{-2}  \label{gAL}
\ee
where $\tilde{F}_1^\pi=\frac{m_N}{m_L}  F_1^\pi$.    What's surprising with this formula is that
\be
 g_A^L\simeq  1
 \ee
 with very little dependence on density near $n_0$ with the density dependence compensating in the two factors\footnote{Even in light nuclei. This result, already observed in \cite{MR91},  has remained resilient since then. I will however have to mention a potentially significant new development by the most recent RIKEN data for the Gamow-Teller transition in $^{100}$Sn I will come to later. It could seriously invalidate this result (\ref{gAL}). It will involve a possible quantum anomaly effect relevant to the IR fixed point structure of the QCD-CD model~\cite{Zwicky}}.  I am assuming that  $g_A^L$ 
 can be taken as the quasiparticle Landau fixed-point quantity. It corresponds to the {\it effective axial coupling constant} for a quasiparticle on the Fermi surface making the {\it superallowed}  quasiproton$\leftrightarrow$quasineutron $\beta$ transition.  What's most remarkable about this relation, apart from its simple form, is that it involves {\it only} the scaling of the dilaton condensate $\Phi$ and the pionic Landau parameter. Furthermore it turns out to connect even to the dialton-limit fixed point (\ref{DLFP})  at much higher density, a feature which will turn out to be relevant for compact stars. 

Let me mention here that there are recent developments in this expression (\ref{gAL}) which contain a lot yet to be clarified. For instance, (\ref{GT}) is  the well-known Goldberger-Treiman relation which when calculated in the Skyrme soliton model, relates $g_A$ to the coefficient of the Skyrme quartic term which is responsible for the stabilization of the soliton. It has recently been found  to be encoded in the kinetic energy term of the generalized Grassmannian model encapsulating  ``all $a=2$ results" of HLS~\cite{yamawaki}. It is not a parameter  but an essential ingredient of the HLS theory ``begging" to be connected to, say,  holographic dual approaches. Another hidden source of significance is the intervention of the dilaton property in $\Phi$ related to the emergent parity-doubling symmetry~\cite{HKL} in the nucleon in medium and the Landau parameter $F_1$ in the $\omega$ channel  which leads to properties manifesting above the density $n_{HQC}$ in dense compact stars.  It will lead to the connection between $g_A^L$ and pseudo-conformality in compact-star matter.

$\bullet$ {\bf  Quenching of $g_A$ and Trace Anomaly:}

Let me explain what the quantity  (\ref{gAL}) physically corresponds to in finite nuclei. 

 If one considers $f_0(500)$ taken as the dilaton $\hat{\sigma}$ as ``heavy" as the HLS mesons, then $\hat{\sigma}$ should figure as an explicitly relevant field in G$n$EFT. Then following the GD scheme, the nucleon axial-current  will have the anomalous dimension $\beta_{\rm IR}^\prime$  appearing in Callan-Szymanski RG flow~\cite{GD}
\be
J_5^{a\mu}=q_{ssb}g_A\bar{\psi}\gamma^\mu\gamma_5 \frac{\tau^a}{2}\psi +\cdots
\ee 
where $\cdots$ contain fluctuating dilaton fields that do not figure at the leading chiral-scale order and 
\be
q_{\rm ssb} = c_A+(1-c_A)(\frac{f_{\hat\chi}^\ast}{f_{\hat\chi}})^{\beta^\prime}\label{qssb}
\ee
where $\ast$ represents the density dependence and $1-q_{ssb}\neq 0$  the scale (or conformal) anomaly-inducement factor to $g_A$.  $c_A$ is an incalculable potentially density-dependent parameter. I will identify $\delta q_{\rm ssb}=(1-q_{\rm ssb})$ as a ``fundamental renormalization (FR)" of $g_A$ in the G$n$EFT framework, not {\it included in pure nuclear correlation effects}.

Now there are two questions raised: The first  is: Is  there indication for $q_{\rm ssb}$ in nature? The answer is yes if one takes the most recent experimental result on the supperallowed Gamow-Teller transition in $^{100}$Sn nucleus. If conformed, this would have a serious implication in going beyond the Standard Model as I discussed in \cite{D-beta}; the second is how this issue has a connection with the notion of pseudo-conformality  in dense compact-star matter?  I must admit I cannot give a simple and fully satisfactory answer, even though I have written a series of notes on it. Here let me give a brief answer which might come close to the answer. 

The most important observation is that Eq.({\ref{gAL}) anchored on Landau(-Migdal) Fermi liquid can be closely mapped to the ``Extreme Single Particle Shell Model (EPSM)" in doubly-magic closed shell nuclei. The nucleus $^{100}$Sn which has the proton and neutron shells completely filled at 50/50 has the advantage of being a heavy nucleus (near nuclear matter density $n_0$) with the doubly magic shells. Its superallowed GT transition from the $^{100}$Sn state (parent) to the lowest single neutron-quasiparticle-proton-quasihole (daughter) state in $^{100}$In  is taken to come closest to the Fermi-liquid fixed point structure on the Fermi surface at the $\bar{N}\to \infty$ limit.  I present this as my conjecture.

This transition has been measured both at GSI~\cite{GSI} and more recently at RIKEN~\cite{RIKEN}. The two results turn out to come out drastically differently. 
As for the GSI data, what amounts to measuring $\sim 95\%$ of the daughter state in $^{100}$In  leads to
\be
(g^{\rm eff}_A)_{\rm GSI}\approx  0.96.\label{gAGSI}
\ee 
I take this as $\delta q_{\rm ssb}^{\rm GSI}\equiv (1-q^{\rm GSI}_{\rm ssb})\approx 0$, that is, no appreciable anomaly-induced quenching (AIQ in short) if any\footnote{A possible caveat to this is given in \cite{D-beta}. The missing $\sim 5\%$ taken into account in the error-quoted GSI value is -- in the caveat -- in the class of uncertainty in taking (\ref{gAL}) equal to the pure shell-model estimate.}. 

On the contrary, the RIKEN result, claimed to be ``improved" with reduced error bars, comes out quite differently:
\be
(g^{\rm eff}_A)^{\rm RIKEN} \approx (0.74-0.88)\label{gARIKEN}
\ee 
giving 
\be
\delta q_{\rm ssb}^{\rm RIKEN}\approx 0.13 - 042 > 0. \label{AIQ}
\ee
I consider this as $(10-40)\%$  ``anomaly-induced quenching (AIQ)"  of $g_A$.  If re-confirmed,  it would make a drastic change in standard nuclear physics as discussed in \cite{D-beta}.  One can see from (\ref{qssb}) that there will involve a large number of (unknown) parameters that cannot at present be calculable from first principles~\cite{sHLS}. Their possible intervention at the leading chiral-scale order could perhaps be accounted for in $\chi$EFT$_\pi$ at higher loop expansion in  low-density nuclear processes given that the dilaton is massive enough to be integrated out. But various nuclear weak processes measured accurately must necessarily be affected by the leading-order AIQ by the fundamental effect (\ref{AIQ}) even at low density.  That would include the weak processes of mass number $A=3$~\cite{wiringa}\footnote{This result in the Monte Carlo calculation for the $A=3$ nucleus can be taken as a {\it } full many-body calculation including the chiral-symmetry power corrections for the many-body exchange axial-current to N$^2$LO relative to the leading single-particle Gamow-Teller (chiral-filtered-out) operator with the free-space coupling constant $g_A=1.267$. It therefore essentially excludes $O(10-40\%)$ AIQ.}.  Now look at (\ref{qssb}) in the QCD-CD scenario with the IR fixed point which argues that 
\be
\beta^\prime_{\rm IR}=0.\label{0betaprime} 
\ee
Then one gets $1-q_{ssb}=0$ which is consistent with the simple shell-model result $g_A^{\rm eff}\approx 1$ in light nuclei $A\lsim 50$~\cite{review-gA}. It is also highly consistent with certain precision (chiral-filtered) nuclear axial-charge transitions, such as  $0^+\leftrightarrow 0^-, \Delta T=1$~\cite{D-beta}, and some well-controlled pion-nuclear interactions (\`a la Goldberger-Treiman relation) in $\chi$EFT$_\pi$. If on the other hand the RIKEN data turns out to be correct, then (\ref{qssb}) cannot be correct, that is, the idea of an IR fixed point in QCD for $N_f\lsim 3$ must be invalid.  Equally seriously, certain un-calculable corrections, i.e., log terms involving the s-quark mass,  to the (vanishing) anomalous dimension in $q_{ssb}$ predicted by \cite{Zwicky} could intervene, impacting  non-trivially the structure of $\beta^\prime$ {\it exposed in nuclear medium invisible at zero density}.  

$\bullet$ {{\bf $g_A^{\rm eff}=1$ to Pseudo-Conformality} }

The key argument presented on the persistent quasi-particle $g_A^{\rm eff}=1$ from low to high density was anchored on mapping Eq.~(\ref{gAL}) of the Fermi-liquid fixed point formula to the ESPM in doubly magic-shell nuclei. The former is implemented with certain topological structure and hidden symmetry degrees of freedom and is valid in the limit that $\bar{N}\to \infty$.  The latter is valid in the limit that particle-hole loop corrections in shell-model, say, in the ESPM are ignorable. Both involve approximations that may be questionable. Here I want to comment on how the mapping could be improved on or made more realistic. 

A recent development in condensed matter physics on how to calculate $1/\bar{N}$ corrections suggests how the above issue could be addressed in the present framework of G$n$EFT. The strategy is to rely on nonlinear bosonization of the Fermi surfaces by the method of coadoint  orbits~\cite{coadjoint}. In condensed matter, the focus is on the fluctuations of the Fermi  surfaces of the electronic systems, but here the problem involves, in addition to baryonic Fermi surfaces, the coupling of hidden symmetry degrees of freedom, so it is clearly more complicated. Given the complexity involved, it would require a lot more detailed work. Nonetheless even the extremely simplified  preliminary result obtained assuming it is more or less correct is quite interesting and informative in its form.  The resultant  $O(1/\bar{N})$ correction comes out to be given in terms of the Landau parameters $\bar{F}_1^\pi$, $\bar{F}_1^\omega$ and the in-medium dilaton condensate $\la\hat{\chi}\ra^\ast$. It is found to be exceedingly small at $n\approx n_0$~\cite{shao}
\be
\delta g_A < 2 \times 10^{-4} g_A^L 
\ee
and is more or less density-independent. What seems to be in action here is an interplay of the (pseudo-)Goldstone bosons $\pi$ and $\hat{\chi}$ with the   BR-scaling $\omega$ meson via the Landau parameter $\tilde{F}_1^\omega$. This mechanism is most likely related, albeit indirectly, to  what happens to the nucleon mass in the half-skyrmion phase at high density due to a similar interplay going toward the parity-doubled chiral-invariant  nucleon mass $m_0\sim f_{\hat{\chi}}$. It is weakly dependent on density, similar to the sound speed in compact stars $v^2_{pcs}/c^2 \simeq 1/3$. This leads to  the notion of an emergent pseudo-conformality at densities above the putative hadron-quark continuity density at or around the skyrmion-to-half-skyrmion transition observed in Fig.~(\ref{scale_inv}).    
\subsubsection*{\it Pseudo-Conformal Sound Speed at High Density}
To simplify the discussion, let me reduce the Lagrangian \LagnEFT for G$n$EFT to the minimum and do the mean-field approximation in the sense of doing the Fermi-liquid fixed point approximation as in \cite{paengetal}. Following the strategy presented in \cite{friman-rho} for transcribing the simplified scale-chiral effective Lagrangian to the Shankar-Polchinski form with (1) the BR scaling,  (2) the topology change at $n_{1/2}$ and (3) the isovector-vector ($\rho$) mesons decoupled in moving toward the DPFP, one obtains
\footnote{Unless otherwise stated, I will work in the chiral limit. There are possible corrections involving the strange-quark mass term to be yet sorted out in applying QCD-CD model~\cite{Zwicky}.}  the (in-medium) VEV of the trace of the energy-momentum tensor $\theta_\mu^\mu$
\begin{eqnarray}
\langle \theta^\mu_\mu \rangle
&=& \langle \theta^{00} \rangle - \sum_i \langle \theta^{ii}\ra = \epsilon - 3 P\nonumber\\
& =& 4V(\langle \hat\chi \rangle) - \langle \hat\chi \rangle \left. \frac{\partial V(\hat\chi)}{\partial \hat\chi} \right|_{\hat\chi = \langle \hat\chi \rangle} \label{TEMT1}
\ee
where $V(\langle \hat\chi \rangle)$ is the dilaton potential. This comes out to be what one gets by doing the mean-field calculation~\cite{paengetal}. This shows that the Fermi surface does not spoil scale symmetry even with the surface coupled to the vector fluctuations in nonlinear bosonizaton in coadjoint orbits. In fact we will arrive at the conclusion that strongly-correlated hadronic interactions do not modify the dilaton potential. This feature accounts for the emergent scale symmetry in compact-star matter.

$\bullet$ {{\bf Emergent parity doubling and $v_{\rm pcs}$ }

Let's now consider how $\langle \hat\chi \rangle$ would behave in the deep half-skyrmion phase.  This issue was mentioned above in connection with the emergence of parity-doubling. 

There have been extensive discussions in the literature on the possible role of parity doubling in hadronic spectrum in dense matter. A recent review in  both normal nuclear matter and dense matter is in \cite{harada-pd}. One could treat the parity doubling either as intrinsic in gauge theory or emergent in QCD.  In the problem I am addressing I will take the notion that it is an emergent manifestation of conformality at high density~\cite{HKL,sasaki-pd}. As detailed in these articles, in half-skyrmion phase at high density, the interplay between the diltaon $\hat{\chi}$ and the vector meson $\omega$ after {\it the charged vector mesons get decoupled} forces the VEV of the $\hat{\chi}$ field tend to go to density-independent $m_0$ in the nucleon mass which renders the  VEV of the trace of energy-momentum tensor $\theta_\mu^\mu$ independent of density in the range of density relevant to compact stars. This behavior resembles closely what happens to $g_A^L$ as the dilaton-limit-fixed-point (DLFP) is approached. This can get manifested precociously in the sound velocity in the pseudo-conformal form~\cite{MR-towards}. How to take fully into account the Fermi surface fluctuations including the hidden symmetry degrees of freedom taken into account in coadjoint orbits in the $g_A^L$  has not been properly analyzed, so I can say little quantitatively on possible deviations from $v^2_{s}/c^2=1/3$ at densities $\gsim 3 n_0$.  However the crucial point in the formalism is that
$\langle \theta^\mu_\mu \rangle$ may not be zero but its derivative with respect to density could be zero  because of the density-independent $m_0$ to which $\la\hat{\chi}\ra^\ast$ tends to~\cite{paengetal}.
Thus
\be
\frac{\partial}{\partial n} \la\theta_\mu^\mu\ra=\frac{\partial \epsilon (n)}{\partial n} \Big(1-3\frac{v_s^2}{c^2}\Big)=0\label{derivTEMT}
\ee
with $v_s^2/c^2=\frac{\partial P(n)}{\partial n}/\frac{\partial \epsilon(n)}{\partial n}$. Now taking that there are no Lee-Wick states in the density range involved, i.e., $\frac{\partial \epsilon (n)}{\partial n}\neq 0$, one arrives at the pseudo-conformal sound speed
\be
v^2_{pcs}/c^2\approx 1/3.\label{pcs}
\ee

At what density the pseudo-conformal sound speed manifests itself cannot be pinned down precisely by the theory but it must appear when the charged vector mesons decouple from pions as one approaches the dilaton-limit fixed point. This feature could be analyzed in the Grassmannian model for  the $\rho$ meson where such a decoupling takes place~\cite{yamawaki}.

It has been seen where $v^2_{pcs}/c^2\approx 1/3$ appears depends on where the half-skyrmion phase sets in and where the $\rho-\pi$ decoupling takes place.  Figure~\ref{PCS} for $n_{1/2}=2.0 n_0$ exemplifies the different  structure of the pseudo-conformal velocity within the given range. In the absence of the $\rho$ decoupling from $\pi$ and $\omega$ at  or near $n \gsim  n_{1/2}$, it is found ~\cite{kuoetal} that the $v_s$ behaves similarly to the well-known nuclear EFT \`a la APR~\cite{APR} and also what seems to be found in the recent Bayesian inference analyses~\cite{weise}. This is seen in the left panel of Fig.~\ref{PCS}. However if the matter is in the half-skyrmion phase with the $\rho$ mesons starting to decouple from the pions and  the $\omega$ and other hadrons moving toward near the DLFP,  then a typical pseudo-conformal $v^2_{s}/c^2\equiv v^2_{pcs}/c^2\sim 1/3$ is predicted for $n_{1/2} \approx (2.0 -3.5)n_0$ as shown in the right panel. 
\begin{figure}[h]
\begin{center}
\includegraphics[height=3.1cm]{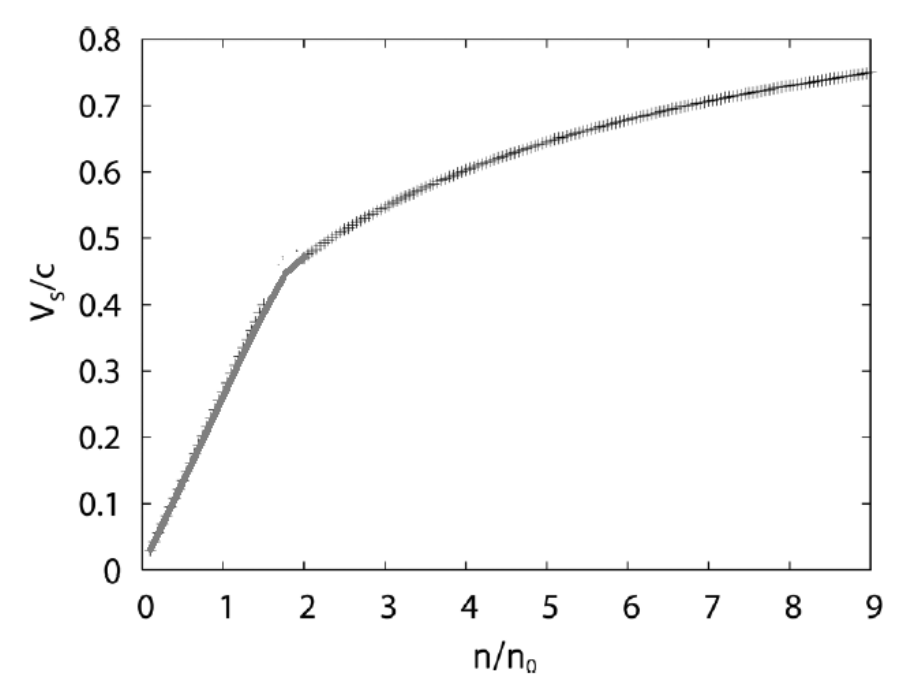}\includegraphics[height=2.9cm]{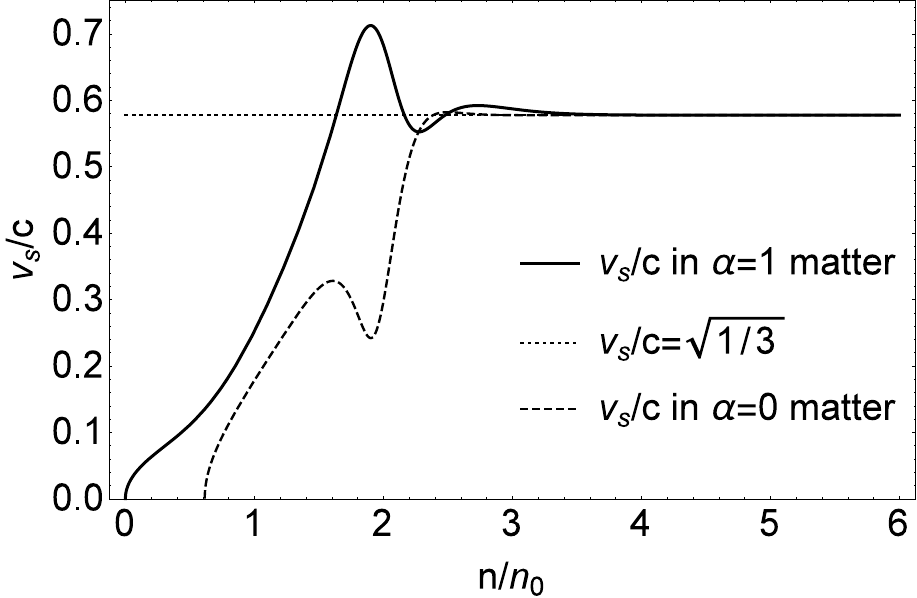}
\caption{Sound speed vs. density for $n_{1/2}=2.0 n_0$: Left panel describes the case where quark-gluon degrees of freedom enter at as low a density as $\sim 6n_0$, assumed to be the core density of massive stars;  Right panel, they figure at much higher density above the central density of compact stars;  $\alpha =\frac{N-P}{N+P}$ where $N(P)$ is the number of neutrons (protons). }\label{PCS}
 \end{center}
\end{figure}
In the right panel, the strength of the peak grows at the half-skyrmion density as  $n_{1/2}$ is moved upwards. For $n_{1/2}\simeq 4.0n_c$, the peak even goes outside of the unitarity bound making it unphysical. In this formalism the maximum star mass for this $n_{1/2}$ is then predicted to be $M_{max}\approx 2.4 M_\odot$.  Whether or not this prediction is at odds with presently available data is not at all clear at the moment. (For amusement let me  stress here that no imaginations are lacking in what theorists are predicting for what nature might have in store. Just look at Fig.~(5.5) in \cite{Komoltsev}. The possibilities would completely fill up the blank space in the figure on the left panel of Fig.~(\ref{PCS}).  I return to \cite{Komoltsev} below for a possible future development.)

An important feature that is clarified in the Grassmannian model for the $\rho$ mesons~\cite{yamawaki} is that the $\rho$ does decouple before reaching the dilaton limit fixed point (DLFP) while the $\omega$ does not, closely locked with the property of the dilaton as seen before, thus differing  from the vector manifestation fixed point (VMFP)~\cite{HLS,yamawaki}. For Fig.~\ref{PCS} this is simulated by the way the $\rho$ and $\omega$ couplings differ at the half-skyrmion phase causing breakdown of the global $U(2)$ symmetry for the vector mesons~\cite{paengetal}. The precise form of the density dependences is clearly not crucial. What maters is where the decouplings take place. 

What's given in Fig.~\ref{PCS} is the result of applying $V_{lowk}$RG strategy with the scaling properties of both before and after $n_{1/2}$ inserted into the EoS. What was done in \cite{paengetal} is that the energy density for $n \geq n_{1/2}$ was parameterized with due account of scale invariance given by Fig.~(\ref{scale_inv}).  This calculation reproduced exactly the $V_{lowk}$RG result. It should of course be admitted that possible effects of higher-order $1/\bar{N}$ corrections \`a la \cite{coadjoint} should  be studied. This might cause certain fluctuating deviations from 1/3 until it reaches the truly conformal limit. 

\section*{Results and Comments}
To conclude, this note presents a parallel between the quenched $g_A^{\rm eff}=g_A^L\approx 1$ from light nuclei to nuclear matter to dense matter and the sound velocity in compact stars from the half-skyrmion phase density  $n_{1/2}\approx 2 n_0$ to the interior density of compact stars $n_{stars}\approx (5-7) n_0$. The parallel is made in terms of a pseudo-conformal symmetry emergent in the coarse-grained description of nuclear matter anchored on a possible existence of an IR fixed point in QCD for $N_f\lsim 3$. This scenario would be ruled out if the RIKEN data for $^{100}$Sn Gamow-Teller superallowed transition were solidly reconfirmed. This RIKEN scenario would also drastically revamp nuclear weak and electromagnetic processes,  pion-nuclear interactions and other so-far well-established nuclear processes.

Let me now give  a few notable results for various astrophysical observables predicted in this G$n$EFT framework. The only parameter in the calculation is the half-skyrmion phase density $n_{1/2} \sim (2.0 - 3.5) n_0$ considered in the model to be the HQC density. The results are slightly dependent on $n_{1/2}$ that I will consider to be acceptable within the formalism largely based on large $N_c$ and $\bar{N}$ considerations. Possible other parameters are fixed on applying the G$n$EFT model slightly above nuclear matter density including $f_{\hat\chi}^\ast\approx f_\pi^\ast$ which should hold for $n\leq 2n_0$. For the range of $n_{1/2}= (2.5 - 3.0)n_0$ the predicted values in pseudo-conformal model (PCM) are (reviewed in \cite{PCMprediction})
\begin{itemize}
\item  {\bf Maximum mass star $M^{\rm max}$}:
\be
{ M^{max} \approx 2.05 M_\odot, 
 \ R_{2.0}\approx 12.8 \ {\rm km},\ 
(n_{central}\approx 5.1 n_0.)}\nonumber
\ee

\item {\bf 1.44  $M_\odot$ star}:
\be
 R_{1.44}\approx 12.8\ {\rm km}.\nonumber
\ee
\end{itemize}
The PCM predictions for some other observables listed in \cite{PCMprediction,MR-towards} are more or less consistent with recent fine  Bayesian analyses~\cite{weise}.  Note further the difference $\Delta R=R_{2.0}-R_{1.4}\approx 0$ in agreement with the data.  As announced the maximum mass comes out to be $\sim 2.4\ M_\odot$ for $n_{1/2}=4 n_0$. However at this crossover density, although other global properties are not drastically different from the lower values of $n_{1/2}$, the sound speed overshoots the causality bound with a more pronounced bump and the pressure greatly exceeds what's indicated in heavy-ion data. It seems to be ruled out in the PCM of G$n$EFT.

A notable difference from the Bayesian analyses of \cite{weise} is seen in the central density of the $M^{max}$.  The PCM predicts roughly a $\sim (20-30)$\% denser maximum density than what's listed in \cite{weise}. This is accounted for by the HQC at near $n_{1/2}$. It is a prediction, e.g., the cusp structure, of the PCM.

The predicted tidal deformability $\Lambda_{\rm 1.4}\sim 550$ seems to come out a bit higher than the quoted value $\Lambda=190^{+300}_{-120}$. In the PCM,  the density at which $\Lambda_{\rm 1.4}$ is measured is $\sim 2.4n_0$,  near $n_{1/2}$. It is close to where  the HQC is to take place.  It is clearly the hardest density regime to access theoretically both bottom-up (EFT) and top-down (perturbative QCD) approaches.   Thus it does not look feasible to pin it down precisely at this stage of theoretical development.

Now a striking difference of the PCM predictions is the sound velocity. The sound velocity concerned here probes the central density $\sim (5-7)n_0$. At this density the PCM velocity $v_{pcs}^2/c^2\approx 1/3$  comes about because $\la \theta_\mu^\mu\ra$ is density-independent since it depends on the invariant mass $m_0$. But  $m_0$ is is not directly related to the``trace anomaly measure" $\Delta=\frac 13 - P(\epsilon)/\epsilon$ until it gets to near the dilaton limit fixed point which must lie at a much higher density. Thus $\Delta$ cannot give an unambiguous constraint for PCM at the central density of the star. In PCM, $\Delta$ reaches 0 at asymptotic density from above rather than from below as in \cite{weise}.  

Finally let me make three remarks, two reiterations and one on future direction: First, the connection between the $g_A^L$ and the PCM sound speed and second, the implication of the RIKEN data on the anomalous dimension $\beta^\prime_{\rm IR}$ and third, the reverse to HQC, say, ``quark-hadron  continuity (QHC)."

First starting with the Goldberger-Treiman relation at large $N_c$ limit in the Skyrme topological model for nucleons and applying the Landau Fermi-liquid fixed point approximation in Fermi-liquid structure of nuclear matter to the chiral Lagrangian implemented with hidden local symmetry and hidden scale symmetry,   the ``effective" axial-current coupling constant in nuclear matter is found to be $g_A^L\approx 1$ valid from light nuclei to nuclear matter and then to the dilaton limit fixed point at high density  where the dilaton condensate approaches $\la\hat{\chi}\ra\to  m_0$, the emergent chiral invariant  nucleon mass where $g_A^{\rm eff}=g_A^{fund}$ goes to exactly 1. This chain of connection is made in the large $N_c$ and large $\bar{N}$ limit. The Goldberger-Treiman relation works fairly well in the matter-free vacuum and is most likely valid in medium. But the corrections to the $1/\bar{N}\to 0$ approximation need still to be worked out.

Second, if the RIKEN data for $g_A^{\rm eff}$ is reconfirmed experimentally, backed by reliable theoretical inputs, then  the AIQ, $\delta q_{\rm ssb}^{\rm RIKEN}$, could be significant. Then as stated the chain of connections given above will be completely broken.  Given that the RIKEN data could, as claimed, to be more reliable than the GSI data, this experiment deserves to be given a high priority for a repeat run in nuclear physics circle.

Third, I have not addressed the reverse, QHC, top-down approach to HQC. It should of course be feasible to go top-down from perturbative QCD  to QHC  with or without phase transitions. In fact there is an active research going on making impressive progress in that direction~\cite{Komoltsev}.  In \cite{PCMspeed}, I discussed some of the PCM predictions compared with the available top-down approach~\cite{Annala} with the cusp structure giving what seems to be required  repulsion followed by attraction in the vicinity of the HQC region. Given that the bottom-up approach can arrive even at the color-favor-locking at high density in QCD in terms of hadronic variables~\cite{CFL} the future development in the top-down direction would be extremely exciting. Personally I would be intrigued how symmetries emergent bottom-up, such as $m_0$, the cusp in $E_{sym}$, etc.  in PCM manifest top-down in QCD variables.

\subsection*{Acknowlegements}

I am deeply grateful for continuous discussions with Hyun Kyu Lee on all issues involved in the PCM during and following the WCU/Hanyang project.


\begin{thebibliography}{99}
\bi{wilkinson} D.H Wilkinson, in {\it Nuclear Physics with Heavy Ions and Mesons}\ (North-Holland, Amsterdam, 1978) eds. by R. Balian, M Rho and G. Ripka.

\bi{FT} S.~Weinberg,
``What is quantum field theory, and what did we think it is?,''
[arXiv:hep-th/9702027 [hep-th]]; “ Effective field theory, past and future,” Int. J. Mod. Phys. A 31, no.06, 1630007 (2016).

\bi{FT-MR} M.~Rho,
``The \textquotedblleft{}Folk Theorem\textquotedblright{} on effective field theory: How does it fare in nuclear physics?,''
J. Korean Phys. Soc. \textbf{71} (2017) no.7, 374-395 doi:10.3938/jkps.71.374 [arXiv:1707.04857 [nucl-th]].

\bi{PL} S.~Weinberg, “Phenomenological Lagrangians,” Physica A 96, no. 1-2, 327 (1979).

\bi{bira-review} 
H.~W.~Hammer, S.~K\"onig and U.~van Kolck,
``Nuclear effective field theory: status and perspectives,''
Rev. Mod. Phys. \textbf{92} (2020) no.2, 025004
doi:10.1103/RevModPhys.92.025004
[arXiv:1906.12122 [nucl-th]].


\bi{Alp-etal} F.~Alp, Y.~Dietz, K.~Hebeler and A.~Schwenk,
``Equation of state and Fermi liquid properties of dense matter based on chiral EFT interactions,''
[arXiv:2504.18259 [nucl-th]].

\bi{multifacet} {\it The Multifaceted Skyrmion: Second Edition}\ (World Scientific Publishing Co. 2017) Ed. by M.  Rho and I.  Zahed.


\bi{park-vento} B.Y. Park and V. Vento, ``Skyrmion approach to finite density and  temperature,"  in \cite{multifacet}.

\bi{Manton} N.S. Manton, {\it SKYRMIONS  A Theory of Nuclei}\ (World Scientific Publishing Co. 2022).

\bi{MR-towards} Y.~L.~Ma and M.~Rho,
``Towards the hadron\textendash{}quark continuity via a topology change in compact stars,''
Prog. Part. Nucl. Phys. \textbf{113} (2020), 103791  [arXiv:1909.05889 [nucl-th]].

\bibitem{HLS}  M.~Bando, T.~Kugo and K.~Yamawaki,
  ``Nonlinear realization and hidden local symmetries,''
  Phys.\ Rept.\  {\bf 164}, 217 (1988);   M.~Harada and K.~Yamawaki,
  ``Hidden local symmetry at loop: A New perspective of composite gauge boson and chiral phase transition,''
  Phys.\ Rept.\  {\bf 381}, 1 (2003).
  
  \bi{yamawaki} K.~Yamawaki,
``Proving $\rho$ meson is a dynamical gauge boson of hidden local symmetry,''
Symmetry \textbf{15}, no. 12, 2209 (2023) [arXiv:2310.09487 [hep-ph]].
 
  \bi{GD} R.~J.~Crewther, ``Genuine dilatons in gauge theories,''
Universe \textbf{6}, no.7, 96 (2020) [arXiv:2003.11259 [hep-ph]]; R.~J.~Crewther and L.~C.~Tunstall,
  ``$\Delta I=1/2$ rule for kaon decays derived from QCD infrared fixed point,''
  Phys.\ Rev.\ D {\bf 91}, no. 3, 034016 (2015).
  

   
  \bi{Zwicky} R.~Zwicky, ``QCD with an infrared fixed point and a dilaton,''
Phys. Rev. D \textbf{110}, no.1, 014048 (2024)
[arXiv:2312.13761 [hep-ph]].  

  \bi{Zwicky-pion}  R.~Zwicky, ``QCD with an infrared fixed point: The pion sector,''
Phys. Rev. D \textbf{109}, no.3, 034009 (2024) [arXiv:2306.06752 [hep-ph]].

\bi{sHLS} Y.~L.~Li, Y.~L.~Ma and M.~Rho,
``Chiral-scale effective theory including a dilatonic meson,''
Phys. Rev. D \textbf{95}, no.11, 114011 (2017) [arXiv:1609.07014 [hep-ph]]; Y.~L.~Li, P.~S.~Wen, Y.~L.~Ma and M.~Rho,
``Scale-chiral effective field theory for nuclear interactions in the Veneziano limit,''
[arXiv:1802.08140 [nucl-th]].  

\bi{MR91} M.~Rho,
``Exchange currents from chiral Lagrangians,''
Phys. Rev. Lett. \textbf{66}, 1275-1278 (1991).

\bi{migdal}  A.B. Migdal, {\it Theory  of Finite Fermi Systems and Applications to Finite Nuclei}\ (Interscience, London, 1967).

\bi{gelmini} G.~Gelmini and B.~Ritzi,
``Chiral effective Lagrangian description of bulk nuclear matter,'' Phys. Lett. B \textbf{357}, 431-434 (1995).

\bi{doubledecimation} G.~E.~Brown and M.~Rho,
``Double decimation and sliding vacua in the nuclear many body system,''
Phys. Rept. \textbf{396}, 1-39 (2004).

\bibitem{matsui} T.~Matsui,
  ``Fermi liquid properties of nuclear matter in a relativistic mean-field Theory,''
  Nucl.\ Phys.\ A {\bf 370}, 365 (1981).

\bi{shankar-polchinski}  R.~Shankar,
 ``Renormalization group approach to interacting fermions,''
  Rev.\ Mod.\ Phys.\  {\bf 66}, 129 (1994); J.~Polchinski,
``Effective field theory and the Fermi surface,''
[arXiv:hep-th/9210046 [hep-th]].

\bi{friman-rho} B.~Friman and M.~Rho,
``From chiral Lagrangians to Landau Fermi liquid theory of nuclear matter,''
Nucl. Phys. A \textbf{606}, 303-319 (1996).

\bi{BRscaling} G.~E.~Brown and M.~Rho,
``Scaling effective Lagrangians in a dense medium,''
Phys. Rev. Lett. \textbf{66}, 2720-2723 (1991).

\bi{lee-park-rho} H.~K.~Lee, B.~Y.~Park and M.~Rho,
``Half-skyrmions, tensor forces and symmetry energy in cold dense matter,''
Phys. Rev. C \textbf{83}, 025206 (2011) [erratum: Phys. Rev. C \textbf{84}, 059902 (2011)].

\bi{senthil} T. Senthil {\it et al.}, ``Deconfined quantum critical points," in \cite{multifacet}.

\bi{cho-monopole} P.~Zhang, K.~Kimm, L.~Zou and Y.~M.~Cho,
``Re-interpretation of Skyrme theory: New topological structures,''
[arXiv:1704.05975 [hep-th]]; Y.~M.~Cho, K.~Kimm, J.~H.~Yoon and P.~Zhang,
``New topological structures of Skyrme theory: Baryon number and monopole number,''
Eur. Phys. J. C \textbf{77}, no.2, 88 (2017).

 
\bi{coadjoint} L.~V.~Delacretaz, Y.~H.~Du, U.~Mehta and D.~T.~Son,
``Nonlinear bosonization of Fermi surfaces: The method of coadjoint orbits,''
Phys. Rev. Res. \textbf{4}, no.3, 033131 (2022)
doi:10.1103/PhysRevResearch.4.033131 
[arXiv:2203.05004 [cond-mat.str-el]].

\bi{paengetal}  W.~G.~Paeng, T.~T.~S.~Kuo, H.~K.~Lee, Y.~L.~Ma and M.~Rho,
``Scale-invariant hidden local symmetry, topology change, and dense baryonic matter. II.,''
Phys. Rev. D \textbf{96}, no.1, 014031 (2017).

\bi{atiyah-manton} B.~Y.~Park, D.~P.~Min, M.~Rho and V.~Vento,
``Atiyah-Manton approach to skyrmion matter,''
Nucl. Phys. A \textbf{707}, 381-398 (2002) [arXiv:nucl-th/0201014 [nucl-th]].

\bi{appelquist}  T.~Appelquist, J.~Ingoldby and M.~Piai,
``Dilaton effective field theory,'' Universe \textbf{9}, no.1, 10 (2023).

\bi{HKL} H.~K.~Lee,
``Parity doubling in dense baryonic matter as an emergent phenomenon and pseudo-conformal phase,''
Symmetry \textbf{16}, no.12, 1598 (2024).

\bi{KDR} K.~Kubodera, J.~Delorme and M.~Rho,
``Axial currents in nuclei,''
Phys. Rev. Lett. \textbf{40}, 755-758 (1978).


\bi{HRW} J.~W.~Holt, M.~Rho and W.~Weise,
``Chiral symmetry and effective field theories for hadronic, nuclear and stellar matter,''
Phys. Rept. \textbf{621}, 2-75 (2016) [arXiv:1411.6681 [nucl-th]].

\bi{NRZ} M.A. Nowak, M. Rho and I.Zahed, {\it Chiral Nuclear Dynamics}\ (World Scientific 1996).

\bi{D-beta} M.~Rho,
``Anomaly-induced quenching of $g_A$ in nuclear matter and impact on search for neutrinoless \ensuremath{\beta}\ensuremath{\beta} decay,''
Symmetry \textbf{15}, no.9, 1648 (2023).

\bi{GSI}  C.B. Henke {\it et al.},
``Superallowed Gamow-Teller decay of the doubly magic nucleus $^{100}$Sn,"
 Nature {\bf 486}, 341 (2012).

\bi{RIKEN} D.~Lubos {\it et al.},
 ``Improved value for the Gamow-Teller strength of the $^{100}$Sn beta decay,''
  Phys.\ Rev.\ Lett.\  {\bf 122}, 222502 (2019).
  
   \bi{wiringa} G.~B.~King, L.~Andreoli, S.~Pastore, M.~Piarulli, R.~Schiavilla, R.~B.~Wiringa, J.~Carlson and S.~Gandolfi,
``Chiral effective field theory calculations of weak transitions in light nuclei,'' Phys. Rev. C \textbf{102}, no.2, 025501 (2020).

\bi{review-gA} J.T.  Suhonen,
``Value of the axial-vector coupling strength in $\beta$  and $\beta\beta$ decays: A Review,"
  Front.\ in Phys.\  {\bf 5}, 55 (2017); 
J.~Engel and J.~Men\`endez,
 ``Status and future of nuclear matrix elements for neutrinoless double-beta decay: A review,''
  Rept.\ Prog.\ Phys.\  {\bf 80}, 046301 (2017).
  

\bi{shao} L.~Q.~Shao and M.~Rho,
``Corrections to Landau Fermi-liquid fixed-point approximation in nonlinear bosonized theory:~An application to $g_A^L$  in nuclei,''
Phys. Rev. C \textbf{110}, no.1, 015204 (2024).

\bi{harada-pd} Y.~K.~Kong, Y.~Kim and M.~Harada,
``Nuclear matter and finite nuclei: Recent studies based on parity doublet model,''
Symmetry \textbf{16}, no.9, 1238 (2024).

\bi{sasaki-pd} W.~G.~Paeng, H.~K.~Lee, M.~Rho and C.~Sasaki,
``Interplay between $\omega$-nucleon interaction and nucleon mass in dense baryonic matter,''
Phys. Rev. D \textbf{88}, 105019 (2013).

\bi{kuoetal} W.~G.~Paeng, T.~T.~S.~Kuo, H.~K.~Lee and M.~Rho,
``Scale-invariant hidden local symmetry, topology change and dense baryonic matter,''
Phys. Rev. C \textbf{93}, no.5, 055203 (2016).

\bi{APR} A.~Akmal, V.~R.~Pandharipande and D.~G.~Ravenhall,
``The equation of state of nucleon matter and neutron star structure,  ''Phys. Rev. C \textbf{58}, 1804-1828 (1998).

\bi{PCMprediction} M. Rho, 
``Dense baryonic matter predicted in {\textquotedblleft}pseudo-conformal model{\textquotedblright},''
Symmetry \textbf{15}, no.6, 1271 (2023).

\bi{weise} L.~Brandes and W.~Weise,
``Implications of latest NICER data for the neutron star equation of state,''
Phys. Rev. D \textbf{111}, no.3, 034005 (2025); 
``Constraints on phase transitions in neutron star matter,''
Symmetry \textbf{16}, no.1, 111 (2024).

\bi{Komoltsev} 
O.~Komoltsev,
``Perturbative QCD reveals the softening of matter in the cores of massive neutron stars,''
[arXiv:2506.06465 [astro-ph.HE]]; E.~Finch, I.~Legred, K.~Chatziioannou, R.~Essick, S.~Han and P.~Landry,
``Unified nonparametric equation-of-state inference from the neutron-star crust to perturbative-QCD densities,''
[arXiv:2505.13691 [nucl-th]].

 \bi{PCMspeed} M.~Rho,
``Pseudo-conformal sound speed in the core of compact stars,''
Symmetry \textbf{14}, no.10, 2154 (2022), 
[arXiv:2209.02327 [nucl-th]].

\bi{Annala} E.~Annala, T.~Gorda, A.~Kurkela, J.~N\"attil\"a and A.~Vuorinen,
``Evidence for quark-matter cores in massive neutron stars,''
Nature Phys. \textbf{16}, no.9, 907-910 (2020).

\bi{CFL} D.~K.~Hong, M.~Rho and I.~Zahed,
``Qualitons at high density,''
Phys. Lett. B \textbf{468}, 261-269 (1999)
[arXiv:hep-ph/9906551 [hep-ph]].
%
\end{thebibliography}
\end{document}